\documentclass[UTF8,letterpaper,preprint,prl,aps,superscriptaddress,floatfix]{revtex4}
\usepackage[latin1]{inputenc}
\usepackage{bm}
\usepackage{multirow,amssymb,amsbsy,amsmath}
\usepackage{graphicx}
\usepackage{verbatim}
\usepackage{color}

\makeatletter
\usepackage{pifont}
\makeatother

\usepackage{graphicx}
\usepackage{dcolumn}
\usepackage{bm}
\usepackage{ifthen}
\usepackage{booktabs}
\usepackage{nicefrac}
\usepackage{float}
\usepackage{bbm}
\usepackage{amsmath}

\newcommand{\ket}[1]{\ensuremath{\left|#1\right\rangle}}

\newcommand{\abs}[1]{\ensuremath{\left| #1 \right|}}

\begin{document}

\title{Experimental demonstration of secure quantum remote sensing}

\author{Peng Yin}
\affiliation{CAS Key Laboratory of Quantum Information, University of Science and Technology of China, Hefei 230026, People's Republic of China}
\affiliation{CAS Center For Excellence in Quantum Information and Quantum Physics, University of Science and Technology of China, Hefei, Anhui 230026, China}

\author{Yuki Takeuchi}
\affiliation{NTT Communication Science Laboratories, NTT Corporation, 3-1 Morinosato-Wakamiya, Atsugi, Kanagawa 243-0198, Japan}

\author{Wen-Hao Zhang}
\affiliation{CAS Key Laboratory of Quantum Information, University of Science and Technology of China, Hefei 230026, People's Republic of China}
\affiliation{CAS Center For Excellence in Quantum Information and Quantum Physics, University of Science and Technology of China, Hefei, Anhui 230026, China}

\author{Zhen-Qiang Yin}
\affiliation{CAS Key Laboratory of Quantum Information, University of Science and Technology of China, Hefei 230026, People's Republic of China}
\affiliation{CAS Center For Excellence in Quantum Information and Quantum Physics, University of Science and Technology of China, Hefei, Anhui 230026, China}

\author{Yuichiro Matsuzaki}
\email{email:matsuzaki.yuichiro@aist.go.jp}
\affiliation{National Institute of Advanced Industrial Science and Technology, Tsukuba, Ibaraki 305-8565, Japan}

\author{Xing-Xiang Peng}
\affiliation{CAS Key Laboratory of Quantum Information, University of Science and Technology of China, Hefei 230026, People's Republic of China}
\affiliation{CAS Center For Excellence in Quantum Information and Quantum Physics, University of Science and Technology of China, Hefei, Anhui 230026, China}

\author{Xiao-Ye Xu}
\affiliation{CAS Key Laboratory of Quantum Information, University of Science and Technology of China, Hefei 230026, People's Republic of China}
\affiliation{CAS Center For Excellence in Quantum Information and Quantum Physics, University of Science and Technology of China, Hefei, Anhui 230026, China}

\author{Jin-Shi Xu}
\affiliation{CAS Key Laboratory of Quantum Information, University of Science and Technology of China, Hefei 230026, People's Republic of China}
\affiliation{CAS Center For Excellence in Quantum Information and Quantum Physics, University of Science and Technology of China, Hefei, Anhui 230026, China}

\author{Jian-Shun Tang}
\affiliation{CAS Key Laboratory of Quantum Information, University of Science and Technology of China, Hefei 230026, People's Republic of China}
\affiliation{CAS Center For Excellence in Quantum Information and Quantum Physics, University of Science and Technology of China, Hefei, Anhui 230026, China}

\author{Zong-Quan Zhou}
\affiliation{CAS Key Laboratory of Quantum Information, University of Science and Technology of China, Hefei 230026, People's Republic of China}
\affiliation{CAS Center For Excellence in Quantum Information and Quantum Physics, University of Science and Technology of China, Hefei, Anhui 230026, China}

\author{Geng Chen}
\email{email:chengeng@ustc.edu.cn}
\affiliation{CAS Key Laboratory of Quantum Information, University of Science and Technology of China, Hefei 230026, People's Republic of China}
\affiliation{CAS Center For Excellence in Quantum Information and Quantum Physics, University of Science and Technology of China, Hefei, Anhui 230026, China}

\author{Chuan-Feng Li}
\email{email:cfli@ustc.edu.cn}
\affiliation{CAS Key Laboratory of Quantum Information, University of Science and Technology of China, Hefei 230026, People's Republic of China}
\affiliation{CAS Center For Excellence in Quantum Information and Quantum Physics, University of Science and Technology of China, Hefei, Anhui 230026, China}

\author{Guang-Can Guo}
\affiliation{CAS Key Laboratory of Quantum Information, University of Science and Technology of China, Hefei 230026, People's Republic of China}
\affiliation{CAS Center For Excellence in Quantum Information and Quantum Physics, University of Science and Technology of China, Hefei, Anhui 230026, China}


\begin{abstract}
Quantum metrology aims to enhance the precision of various measurement tasks by taking advantages of quantum properties. In many scenarios, precision is not the sole target; the acquired information must be protected once it is generated in the sensing process. Considering a remote sensing scenario where a local site performs cooperative sensing with a remote site to collect private information at the remote site, the loss of sensing data inevitably causes private information to be revealed. Quantum key distribution is known to be a reliable solution for secure data transmission, however, it fails if an eavesdropper accesses the sensing data generated at a remote site. In this study, we demonstrate that by sharing entanglement between local and remote sites, secure quantum remote sensing can be realized, and the secure level is characterized by asymmetric Fisher information gain. Concretely, only the local site can acquire the estimated parameter accurately with Fisher information approaching 1. In contrast, the accessible Fisher information for an eavesdropper is nearly zero even if he/she obtains the raw sensing data at the remote site. This achievement is primarily due to the nonlocal calibration and steering of the probe state at the remote site. Our results explore one significant advantage of ``quantumness'' and extend the notion of quantum metrology to the security realm.

\end{abstract}
\maketitle

\section{Introduction}
Imposing quantum properties on information science can inspire innovations with unique advantages over classical counterparts. As paradigmatic applications, a quantum computer can efficiently solve problems that are beyond the capabilities of classical computers \cite{Shor,Grover,Harrow,Vandersypen}, and quantum cryptography, such as quantum key distribution (QKD), can realize information-theoretic secure communications \cite{yzq2019,Bennett1,Bennett2,Gisin}. In the past decades, the demands for solutions to various types of measurement tasks have sparked a new realm, i.e., quantum metrology, which focuses on enhancing measurement precision by utilizing quantum resources, such as superposition and entanglement. Quantum metrology has shown the ability to beat the standard quantum limit that constrains the sensitivity of known classical
methods \cite{Giovannetti1,Giovannetti2,Nagata,Mitchell,Wineland,Huelga,Matsuzaki1,Chin}, or even attain the Heisenberg limit in some scenarios \cite{Chen1,Chen2,zlj2015}.

The power of computation, the security of communication, and the precision of measurement are the main topics of quantum information science. Several in-depth studies have revealed that these three topics are not conventionally independent, and interdisciplinary approaches that combine two topics and eventually lead to advantageous approaches. For example, by combining quantum computation and quantum cryptography, blind quantum computing (BQC) enables a local client with computationally weak devices to delegate universal quantum computation to a remote server that has a universal quantum computer, while preventing the remote server from accessing the information of computation \cite{Broadbent,Morimae,Takeuchi,Barz,Greganti}. 
A network of quantum sensors is also becoming an attractive topic in quantum metrology \cite{Eldredge,Komar,Proctor}, and quantum cryptography can help protect sensing information when transmitted between nodes \cite{Giovannetti3,Giovannetti4,Chiribella1,Chiribella2}. Recently, a delegated remote sensing method with built-in security has been proposed \cite{TMMSM19}. In that study, a local site (client) delegates the sensing task to a remote site (server) that possesses a high-precision sensor; however, the server cannot determine the sensing parameter if the server honestly follows the client's instructions. In other words, even if the server collects all existing classical data at the server's site after the delegated remote sensing protocol, the server cannot obtain sufficient information on the sensing parameter. The security of this proposal was confirmed through a purely mathematical derivation of the upper and lower bounds of the measurement uncertainties on the client and server sides, respectively.

In this study, from a practical perspective, we consider a scenario where a local site performs cooperative sensing with a remote site to collect certain private information at the remote site while simultaneously protecting against any information exposure to an eavesdropper, i.e., Eve. This scenario represents many measurement tasks in which the sensing information must be protected once it is generated. One typical application is diagrammed in Fig. 1 (1), where the doctor in a hospital remotely senses the condition of the patient, while keeps the health information private to any third party. In the classical scheme, if the sensor is stolen by an eavesdropper, he/she can in principle recover all data that were measured before, because the classical information is not erasable. 

To solve this problem, utilizing maximally entangled Bell pairs, we experimentally demonstrate a secure quantum remote sensing (SQRS) approach that can protect the sensing information from its birth. The key element of this achievement is the ability of the local site to non-locally calibrate and steer the probe state at the remote site; thus, Eve cannot determine the initial probe state which is required to extract information from the sensing data. To put it rigorously, the security level achieved in our experiment is quantified by the amount of extracted classical Fisher information (CFI) of the local site and Eve. The proposed method extends the notion of quantum metrology to the security realm, and could be a promising secure technique in various sensing tasks.

\medskip
\section{RESULTS}
\begin{figure}[htbp]
		\centering
		\includegraphics[width=4in]{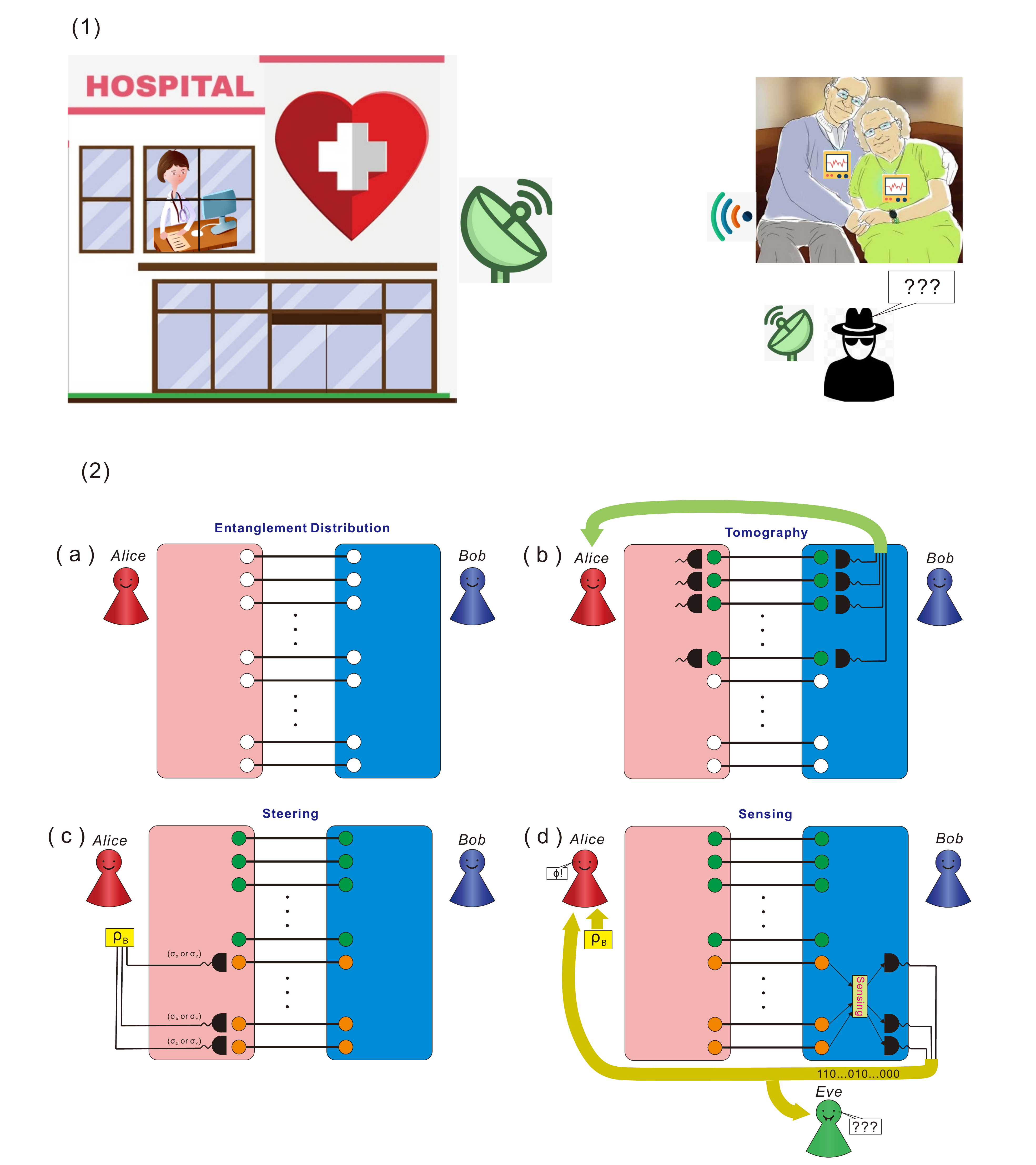} 		
    \caption{\textbf{(1)} \textbf{One typical scenario that requires SQRS.}
    A remote patient needs to check his/her condition using a wearable sensor, and the wearable sensor sends the collected raw data to a local hospital. The patient hopes that a doctor will analyze the data and monitor his/her condition. He/She also does not want anyone else to access his/her personal information. Consequently, it is expected that his/her private information can still be protected even if the classical sensing data in the remote snesor is completely accessed by an eavesdropper.
    \textbf{(2)} \textbf{The steps to perform a SQRS protocol.}
    \textbf{a.} Alice and Bob share photon pairs in Bell states. \textbf{b.} To calibrate the shared state, Alice and Bob perform quantum-state tomography collaboratively on part of the photon pairs, and Bob sends his results to Alice unidirectionally. \textbf{c.} Alice steers Bob's photons by measuring her remaining photons uniformly at random in the Pauli-$X$ or $Y$ basis and maintains the secrecy of the measurement outcomes. As a result, a single-qubit quantum state $\rho_{Bi}$ is prepared on Bob's side where Alice knows the exact form of $\rho_{Bi}$, while Eve possesses no information about the state due to the lack of knowledge about Alice's measurement outcomes. \textbf{d.} With the steered photons, Bob senses the sample and sends Pauli-$Y$ measurement outcomes to Alice. Combining this feedback information with prior knowledge of $\rho_{Bi}$, Alice can determine the parameter precisely while Eve cannot access information about the parameter.}
    \label{TQRS}
\end{figure}

\noindent{\bf Theoretical framework.}
Assume there are cooperative local and remote sites named Alice and Bob in our proposal. Alice is located in a secure area, and any invasive attack is prohibited on her side. Bob is a remote site outside the secure area and may move over time. Bob carries a sensor to perform some sensing tasks, such as health monitoring or prospecting. The readout raw data from the sensor are sent back to Alice, who analyzes the raw data to obtain final results. For security, they should keep the information embedded in the raw data format; thus, any eavesdropper accessing the raw data cannot extract the private information. To achieve this goal, the information must be protected once it is generated in the sensing process. Apparently, this requirement is beyond the capacity of QKD, which was developed to protect information during transmission. We demonstrate that, by using the SQRS protocol, the eavesdropper cannot obtain sufficient information about the sensed parameter even if all classical data generated by Bob's sensor are eavesdropped.

Such SQRS incorporates quantum nonlocality into the standard quantum sensing protocols to achieve a high-level of security. The standard quantum sensing method comprises three procedures, i.e., preparation of the initial qubit state $\rho_{in}$, time evolution of $\rho_{in}$ by interacting with a sample, and projective measurement on the evolved single-qubit state. To perform SQRS, Alice and Bob pre-share entangled photon pairs, which are supposed to be maximally entangled state. Then, the initial probe state preparation is divided into two steps, i.e., nonlocal calibration and steering, which are accomplished cooperatively by Alice and Bob, as shown in Fig. 1 (2b) and (2c).

In detail, considering the unpredictable state deviation that may occur during the entanglement distribution, they first implement a quantum-state tomography~\cite{JKMW01}. After measuring their respective photons, Bob sends his results to Alice unidirectionally. As a result, although Bob can measure his own photon precisely, the overall two-qubit entangled state cannot be learned fully by Bob. In the second step, Alice measures her remaining photons in Pauli-$X$ or $Y$ basis uniformly at random to prepare a single-qubit quantum state $\rho_{Bi}$ on Bob's side, which correspond to Alice's four possible measurement results $Ai$ ($A1 (A3)$ and $A2 (A4)$ correspond to Pauli-$Y$ ($X$) measurements with outcomes 1 and 0, respectively). Here the measurement basis (Pauli-$X$ or $Y$) and corresponding measurement outcome $s_{A}$ (0 or 1) are kept secret by Alice.

 In the last procedure of SQRS protocol, the state $\rho_{Bi}$ evolves by interacting with the sample on Bob's side, and is then measured by Bob in the Pauli-$Y$ basis with measurement outcome $s_{B}$. In other words, $\rho_{Bi}$ is used as the initial probe state rather than $\rho_{in}$. Alice and Bob repeat sensing with $N$ pairs of entangled photons, and Bob sends all measurement outcomes $\{s_{B}^{(j)}\}_{j=1}^N$ to Alice. Here, superscript $(j)$ denotes the $j$th repetition. Alice can then estimate the parameter from the values of $\{s_{A}^{(j)}\oplus s_{B}^{(j)}\oplus 1\}_{j=1}^N$, where the $\oplus$ denotes the addition modulo 2. Note that Eve does not know Alice's measurement bases and corresponding measurement outcomes $\{s_{A}^{(j)}\}_{j=1}^N$; thus, she cannot know the exact form of $\rho_B=\otimes_{j=1}^N\rho_{Bi}^{(j)}$, which prohibits Eve from estimating the parameter. This asymmetric information gain secures the entire protocol, and the sensed parameter can only be obtained by Alice who maintains the secrecy of measurement basis and measurement outcome $s_{A}$. Furthermore, in this protocol, Alice is only required to perform single-qubit measurements. No entangling operation is required for Alice, which is practically important relative to reducing the technological requirements.

\begin{figure}[htbp]
\centering
\includegraphics[width=6in]{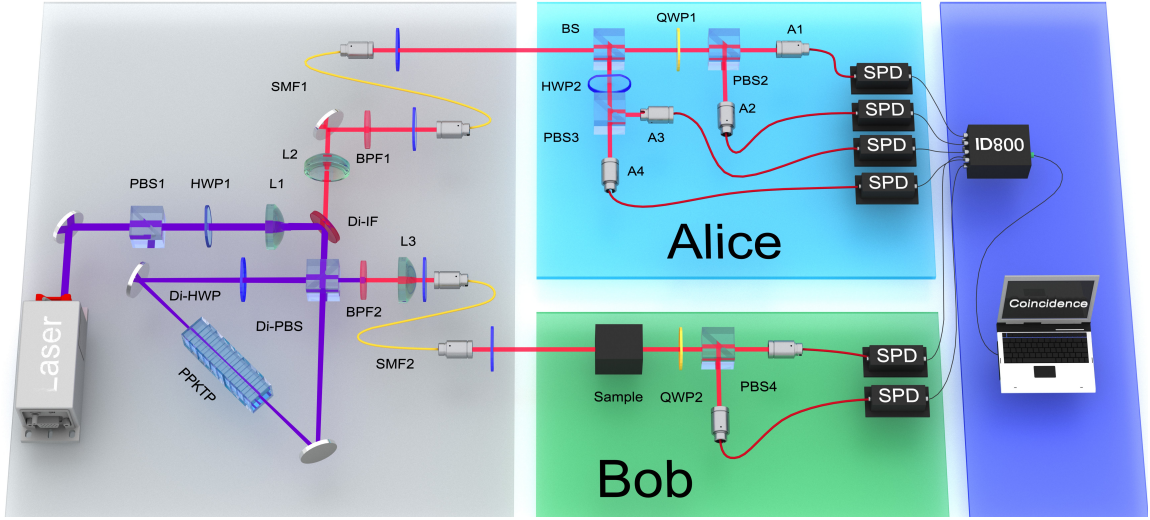}
\caption{\textbf{Experimental setup for SQRS.} Through a type-\uppercase\expandafter{\romannumeral2} spontaneous parametric down conversion process by pumping a periodically poled KTiOPO4 (PPKTP) nonlinear crystal in a Sagnac interferometer (SI), this apparatus generates polarization-entangled photon pairs, which are shared by Alice and Bob to perform SQRS. The shared states are calibrated by a state-tomography process with polarization analyzers composed of polarized beam splitters (PBSs) and corresponding waveplates. A non-polarized beam splitter is placed on Alice's side, in which the two exits correspond to the Pauli-$X$ or $Y$ measurements on her photons; thus, the photons on Bob's side collapse accordingly. Then, Bob implements sensing by coupling his photons with the sample and performs Pauli-$Y$ measurement with his polarization analyzer. The coincidences of Alice and Bob's measurements are recorded by an ID800 (ID Quantique). For each of the four possible combinations of measurement bases and measurement outcomes on Alice's side, the parameter can be estimated by calculating the corresponding projection probability of Bob's measurement. Di: dichroic, BS: beam splitter, IF: interference filter, BPF:band-pass filter; SMF: single mode fiber, HWP: half wave plate, QWP: quarter wave plate, PBS: polarized beam splitter, L: lens, SPD: single photon detector.}
\label{Setup}
\end{figure}

\medskip
{\bf \noindent Experimental apparatus and results.}
The experimental setup to demonstrate SQRS is shown in Fig. 2. The polarization-entangled photon pairs shared by Alice and Bob are generated from a Sagnac interferometer (SI) by pumping a periodically poled KTP (PPKTP) crystal with a single mode 405.4-nm laser (see the Method section for details). Once the entangled photon pairs are shared between Alice and Bob, state tomography is performed to test the quality of the shared state in case of generation and transmission errors. The tomography results show that the fidelity of the shared states to maximally entangled singlets is above 99\%, which guarantees both the security and accuracy of SQRS. Here, $|H\rangle$ and $|V\rangle$ denote the horizontal and vertical polarization states of photons, respectively. In our experiment, $|0\rangle$ and $|1\rangle$ are encoded by $|H\rangle$ and $|V\rangle$, respectively.
Then, Alice randomly implements one of the two projective measurements PM1($\sigma_y$) and PM2($\sigma_x$) on her own photon, each of which will be performed on one of the two exits of a 50/50 non-polarized beam splitter on Alice's side.
Here, $\sigma_y$ and $\sigma_x$ denote the Pauli-$Y$ and $X$ operators, respectively.
Therefore, for each photon there are four possible outcomes $Ai$, as shown in Fig. \ref{Setup}. Note that outcomes $0$ and $1$ correspond to the projections to the $+1$ and $-1$ eigenvalues, respectively. Accordingly, when Alice's outcome is $Ai$, Bob's photon collapses to $\rho_{Bi}$, which is expressed as follows:
\begin{equation}
\label{bob4probe}
\begin{split}
& \rho_{B1}=|R\rangle\langle R|\equiv|H+iV\rangle\langle H+iV|, \\
& \rho_{B2}=|L\rangle\langle L|\equiv|H-iV\rangle\langle H-iV|, \\
& \rho_{B3}=|D\rangle\langle D|\equiv|H+V\rangle\langle H+V|, \\
& \rho_{B4}=|J\rangle\langle J|\equiv|H-V\rangle\langle H-V|. \\
\end{split}
\end{equation}
 Therefore, four possible probe states can be used for sensing on Bob's side; however, the knowledge about the form of each probe state is only accessible by Alice via her measurement basis and measurement outcome $s_{A}$.

\begin{figure}[htbp]
	\begin{minipage}[t]{0.5\linewidth}
		\centering
		\includegraphics[width=3.5in]{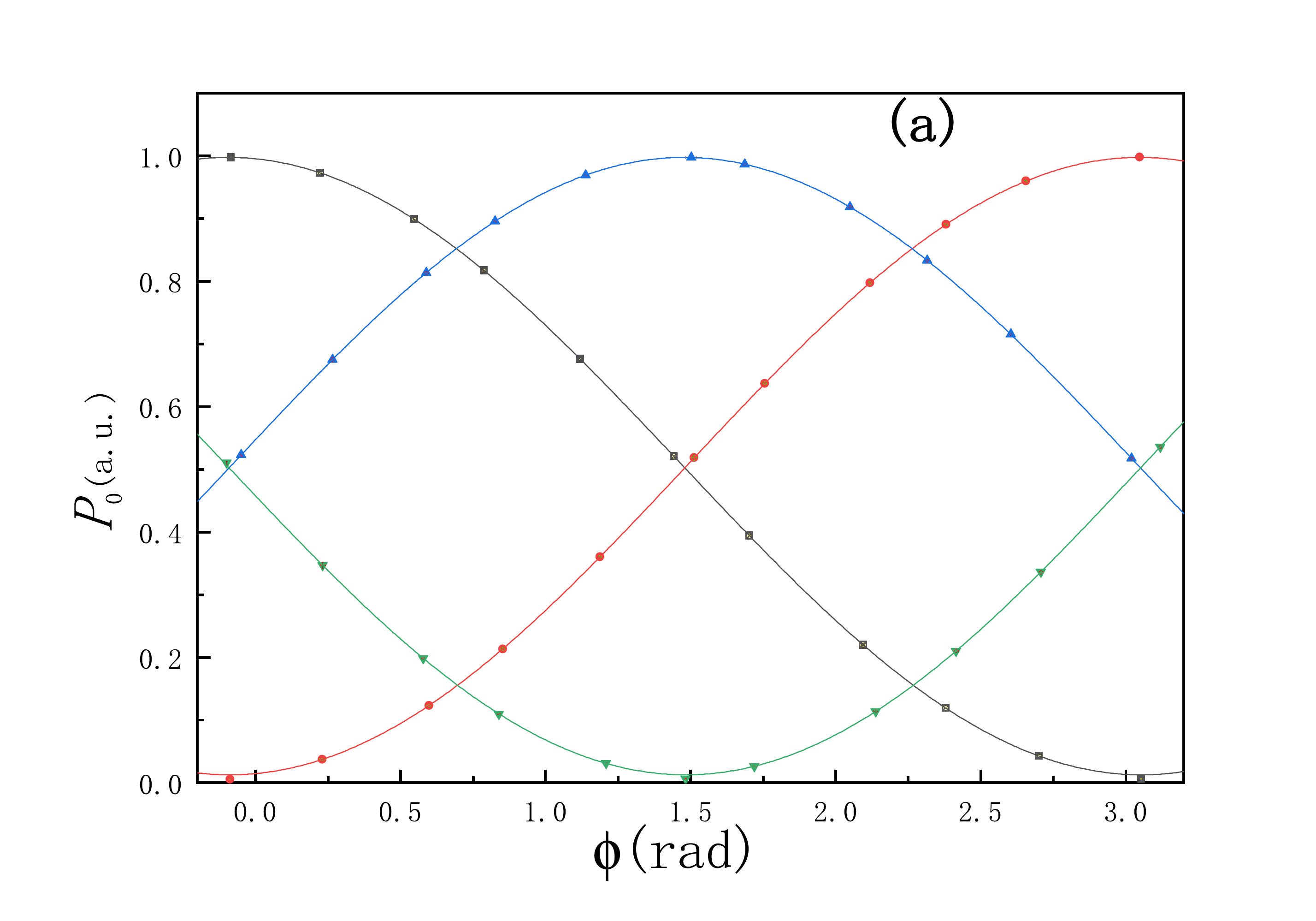} 		
	\end{minipage}%
	\hfill
	\begin{minipage}[t]{0.5\linewidth}
		\centering
		\includegraphics[width=3.5in]{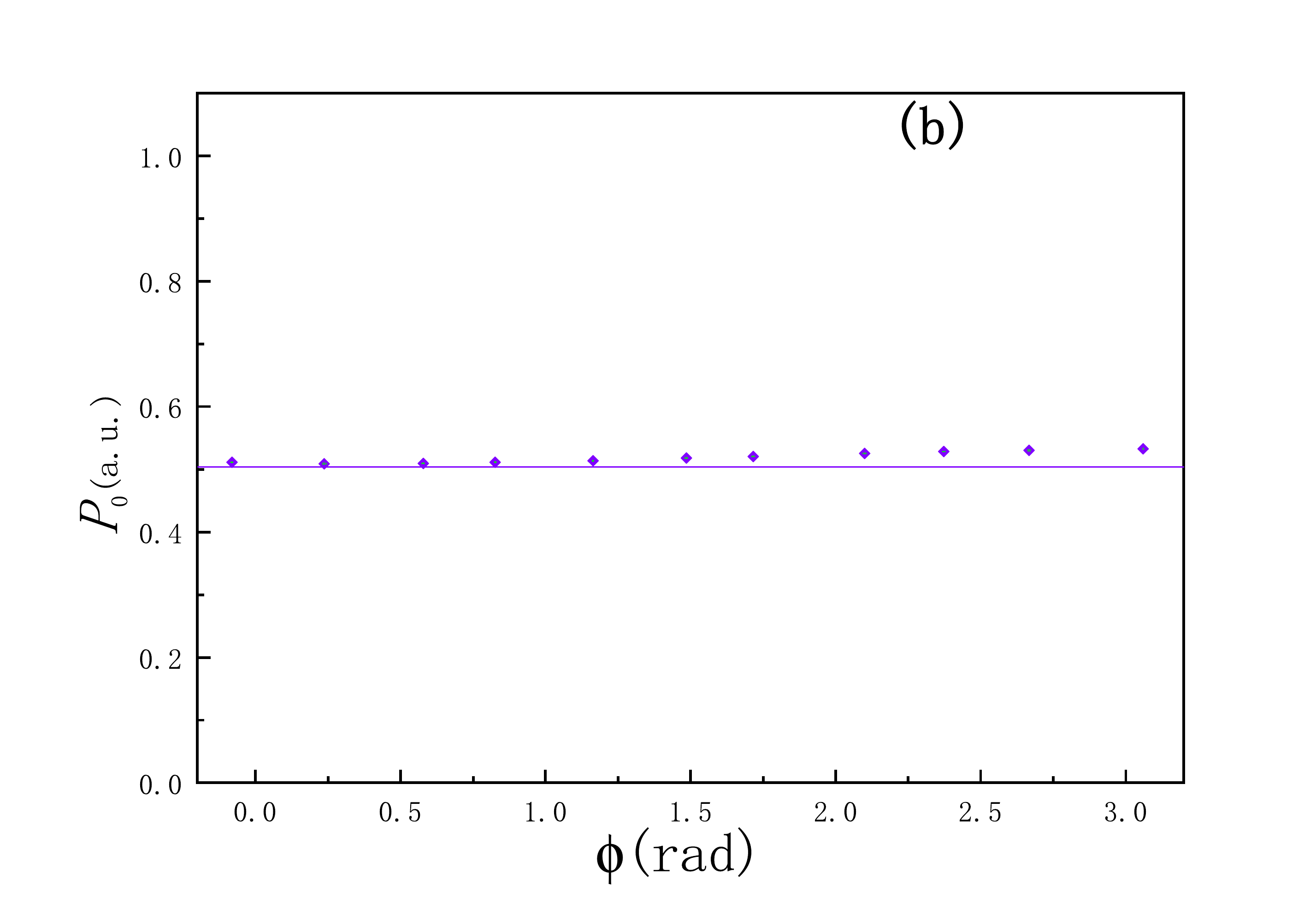} 		
	\end{minipage}
    \flushleft
	\caption{\textbf{Dependence of probabilities on $\phi$.} \textbf{(a)} Black squares, red disks, blue triangles, and green inverted triangles represent the probabilities of obtaining 0 for the four groups of classical sensing data, which correspond to Alice's four measurement outcomes $A1, A2, A3$ and $A4$. \textbf{(b)} Purple diamonds represent the probabilities of Bob obtaining outcome 0. The solid curves that connect these 11 points are their theoretical counterparts. In both subfigures, The error bars are less than the marker size. }
	\label{PtoPhi}
\end{figure}

\begin{figure}[htbp]
	\centering
	\includegraphics[width=5in]{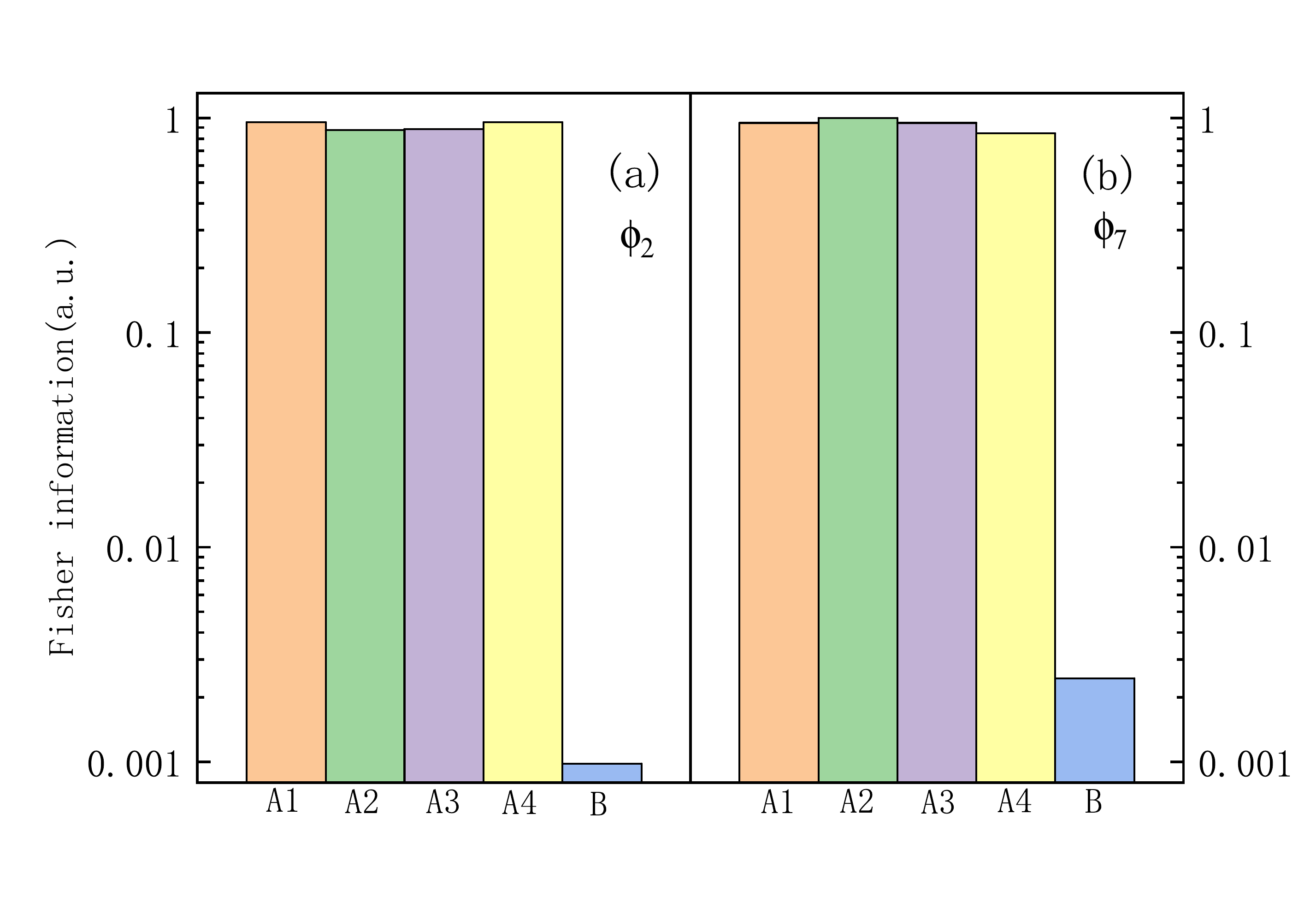}
    \flushleft
	\caption{\textbf{Fisher information of two selected phases.} The measured CFI of (\textbf{a}) $\phi_{2}$ and (\textbf{b}) $\phi_{7}$ in Fig. \ref{PtoPhi} are shown for Alice's four groups of classical sensing data (bars labeled from $A1$ to $A4$), together with the CFI contained in the raw sensing data on Bob's side (bar labeled as B).}
	\label{Fi}
\end{figure}

Here, the sensed parameter is the birefringence of a phase compensation plate (PCP), which can introduce a relative phase $\phi$ between the $|H\rangle$ and $|V\rangle$ components on the transmitted photons. The value of $\phi$ can be controlled by changing the thickness of the PCP via tilting. After the interaction between the probe and sample, Bob implements a projective measurement in the $\sigma_y$-basis and sends the classical sensing data $s_{B}$ to Alice. Alice then classifies the received data into four groups according to $\rho_{Bi}$. 
Then, Alice can estimate the value of $\phi$ according to the projection probability of each group. In our experiment, 11 phases are measured, which are denoted $\phi_{k}$ ($k=0,1,...,10$) and range from 0 to $\pi$. The measured probabilities of the outcome to be 0 for the four groups are shown in Fig. \ref{PtoPhi} (a). For each phase, Alice obtains four estimated values of $\phi_{k}$, which, in principle, should be identical. To be more rigorous, in such calculations, we use the actual form of the probe states obtained from the tomography results (see the Method section for details). In contrast, from the unclassified raw sensing data on Bob's side, the probability of obtaining 0 is nearly a constant with varying $\phi_{k}$, as shown in Fig. \ref{PtoPhi}(b), which prevents Eve from knowing the value of $\phi_{k}$.
Concretely, the secure level can be quantified by the amount of CFI for Alice and Bob. Here,  two measured phases in Fig. \ref{PtoPhi} are selected to calculate the CFI (see the Method section for details), and the results are shown in Fig. \ref{Fi}. A distinct asymmetric information gain is observed, with the CFI on Alice's side approaching 1, while the possible leaked information from Bob's classical sensing data is lower by approximately three orders of magnitude. For perfect singlet states, the probability of obtainning 0 on Bob's side should always be 0.5, which results in a vanishing CFI contained in Bob's classical sensing data. In our experiment, due to the imperfections of states and the nonuniform detection efficiency of single photon detectors (SPDs), this probability varies slightly for different $\phi$ and causes nonvanishing CFI for Bob.
\section{Discussion}
In this experiment, it is reasonable to assume the states are independent and identically distributed (i.i.d.) because our optical experimental setup is stable for the duration of the experiment. As a result, the shared entangled state can be well identified with quantum-state tomography. However, in a long-distance SQRS task, the channel between the cooperative parties may be unstable; thus, the typical parameter (e.g., decoherence rate) would fluctuate temporally. Therefore, the i.i.d assumption becomes invalid. In this case, the random sampling test \cite{NC00,TMMM18F,Zhang} can be adopted because it is impervious to time-variant channel noise \cite{TMMSM19}.

Note that Alice can also implement projective measurement only on one set of basis (i.e., Pauli-Y) to achieve asymmetric information gain; however, there will be vanishing points of CFI in this way. Furthermore, when CFI vanishes for Pauli-X (Pauli-Y) basis, it is almost maximum for Pauli-Y (Pauli-X) basis. Therefore, by implementing projective measurements on Pauli-X and Pauli-Y basis uniformly at random, our experiment can cover the complete range of the sensed parameter without blind point on CFI.

In the proposed protocol, prior entanglement distribution is assumed before the implementation of SQRS; however, this assumption is not absolutely necessary when the eavesdropper cannot access the quantum state in the communication channel, i.e., the eavesdropper only accesses the sensor after sensing is completed. Benefit from this advantage, our protocol is feasible in kinds of  practical scenarios.

In summary, a reliable SQRS is achieved in this work with the aid of entanglement. The results indicate that quantum properties allow precision enhancement and can incorporate security into quantum metrology, which resembles QKD in quantum communication and BQC in quantum computation. In this sense, this work exploits another realistic advantage of utilizing quantum resources, particularly for sensing tasks that requires privacy.

\section{Methods}
\noindent\textbf{Polarization-entangled photon pair source.}
As shown in Fig. \ref{Setup}, a 405.4-nm single mode laser is used to pump the periodically poled KTP (PPKTP) nonlinear crystal placed in a phase-stable SI to produce polarization-entangled photon pairs at 810.8 nm. A dichroic polarized beam splitter (Di-PBS) followed by a dichroic half wave plate (Di-HWP) is used to control polarization of the pump light and generated photon pairs. The lenses positioned before and after the SI focus the pump light and collimate the entangled photons, respectively. A dichroic interference filter (Di-IF) and bandpass filter (BPF) are used to remove the residual pump light. Note that another two BPFs are placed immediately after the exit-side fiber couplers of SMF1 and SMF2; however, for simplicity, these are not plotted in Fig.~\ref{Setup}. Finally, the state of a photon pair that exits the SI can be written as $|singlet\rangle$ in the ideal case.

\noindent\textbf{Calculation of the sensed parameter.}
Once Alice knows the density matrix $\rho$ of the shared states, she can calculate parameter $\phi$ from the probabilities of obtaining 0 for the four groups of classical sensing data. The corresponding probe states on Bob's side before sensing are expressed as follows:
\begin{equation}
\label{bob4realprobe}
\begin{split}
& \rho_{B1}=\frac{Tr_{Alice}[\rho\cdot (|L><L|\otimes I_{2\times 2})]}{Tr[\rho\cdot (|L><L|\otimes I_{2\times 2})]}, \\
& \rho_{B2}=\frac{Tr_{Alice}[\rho\cdot (|R><R|\otimes I_{2\times 2})]}{Tr[\rho\cdot (|R><R|\otimes I_{2\times 2})]}, \\
& \rho_{B3}=\frac{Tr_{Alice}[\rho\cdot (|J><J|\otimes I_{2\times 2})]}{Tr[\rho\cdot (|J><J|\otimes I_{2\times 2})]}, \\
& \rho_{B4}=\frac{Tr_{Alice}[\rho\cdot (|D><D|\otimes I_{2\times 2})]}{Tr[\rho\cdot (|D><D|\otimes I_{2\times 2})]}, \\
\end{split}
\end{equation}
where $Tr_{Alice}[\cdot]$ and $I_{2\times 2}$ represent the partial trace over the qubit possessed by Alice and the two-dimensional identity operator applied to Bob's photon, respectively.
After interaction, the density matrix of Bob's photon evolves to  $\widetilde{\rho_{Bi}}=e^{-i\phi\sigma_z/2}\rho_{Bi}e^{i\phi\sigma_z/2}$, where $i\in {\{1,2,3,4\}}$ corresponds to Alice's outcome $Ai$ and $\sigma_z$ denotes the Pauli-$Z$ operator. Thus, the corresponding probabilities of obtaining 0 are $P^{(i)}(\sigma_y,0)\equiv Tr[\widetilde{\rho_{Bi}}\cdot |R><R|]$. For the ideal singlet state, the dependences of the four probabilities on $\phi$ are relatively simple and can be expressed as follows:
\begin{equation}
\label{bob4realprobability}
\begin{split}
& P_{singlet}^{(1)}(\sigma_y,0)=\frac{1}{2}(1+cos\phi), \\
& P_{singlet}^{(2)}(\sigma_y,0)=\frac{1}{2}(1-cos\phi), \\
& P_{singlet}^{(3)}(\sigma_y,0)=\frac{1}{2}(1+sin\phi), \\
& P_{singlet}^{(4)}(\sigma_y,0)=\frac{1}{2}(1-sin\phi). \\
\end{split}
\end{equation}
With these equations, the value of $\phi$ can be calculated directly from the measured probabilities. However, for the actual shared state used in our experiment, the $\phi$-dependences of the four probabilities are complicated, even though the fidelity between the shared state and the ideal singlet state is greater than 0.99. Then the estimated parameter is taken as the value of $\phi$ that minimizes $|P^{(i)}(\sigma_y,0)-P^{(i)}_{exp}|$, where $P^{(i)}_{exp}$ represents the probability of Bob's measurement outcome to be 0 while Alice's outcome to be $Ai$. For each selected value of $\phi$, Bob's probability of obtaining 0 is $P^{(Bob)}_{exp}=N_{B0}/(N_{B0}+N_{B1})$, where $N_{B0}$ and $N_{B1}$ represent the rate of 0 and 1 in Bob's unclassified raw sensing data. Due to the lack of a strategy to obtain the value of $\phi$ from Bob's unclassified raw sensing data, the corresponding phase to each $P^{(Bob)}_{exp}$ in Fig. 4 is selected to be the average of Alice's four measured phases. The corresponding theoretical line is the average of Alice's four theoretical lines, i.e., $P^{(Bob)}(\sigma_y,0)=\frac{1}{4}\sum_{i=1}^{4}P^{(i)}(\sigma_y,0)$.


\begin{figure}[htbp]
	\centering
	\includegraphics[width=5in]{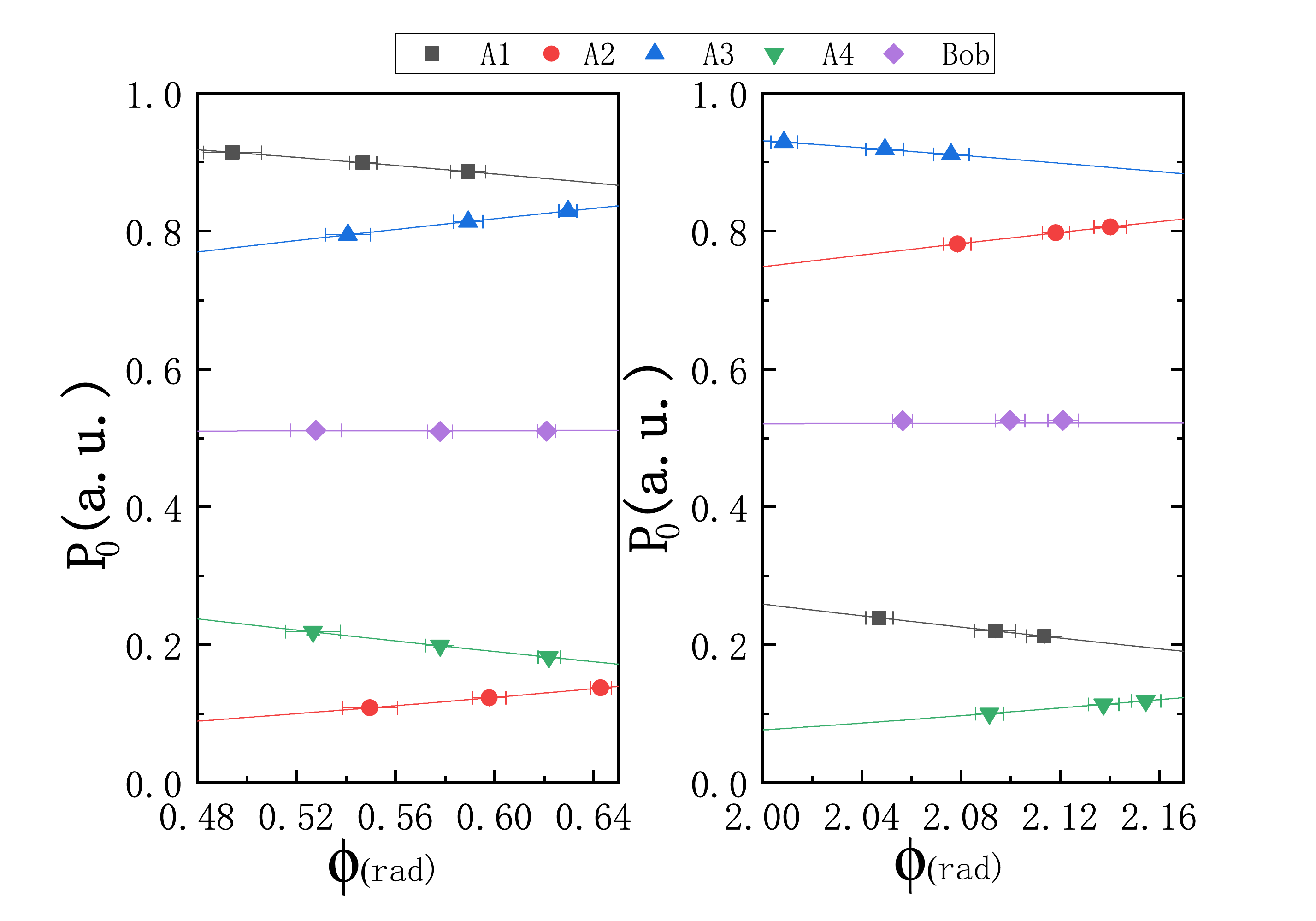}
    \caption{\textbf{Measurement of CFI for (a) $\phi_{2}$ and (b) $\phi_{7}$.} We selected the centering points on each line to calculate the CFI. The two neighboring points are used to approximate the derivatives at the centering points. Thus, the CFI can be calculated from Eq.~(\ref{cfi}).}
	\label{Fip}
\end{figure}

\noindent\textbf{Calculation of CFI.}
To demonstrate the security of the proposed protocol, we used CFI to characterize the asymmetric information gain about the estimated parameter between Alice and a third party, i.e., Eve, who can acquire Bob's classical sensing data. For a Bernoulli distribution, the CFI can be calculated as follows:
\begin{equation}
\label{cfi}
\centering
F_{p}=\frac{1}{P^{(i)}_{exp}}\left(\frac{\partial P^{(i)}_{exp}}{\partial \phi}\right)^{2}+\frac{1}{1-P^{(i)}_{exp}}\left[\frac{\partial\left(1-P^{(i)}_{exp}\right)}{\partial \phi}\right]^{2} =\frac{{\left(\frac{\partial P^{(i)}_{exp}}{\partial \phi}\right)}^2}{P^{(i)}_{exp}\left(1-P^{(i)}_{exp}\right)}
\end{equation}
 The derivative at the centering point can be measured by fitting the slope with two other nearby points (Fig. \ref{Fip}). By inputting the corresponding experimental value into Eq. (\ref{cfi}), we can obtain CFI for (a) $\phi_{2}$ and (b) $\phi_{7}$, as shown in Fig. \ref{Fi}.

$\textbf{Acknowledgments}$ .
 This work was supported by the National Key Research and Development Program of China (Nos. 2017YFA0304100, 2016YFA0302700), Leading Initiative for Excellent Young Researchers MEXT Japan, MEXT KAKENHI (Grant No. 15H05870), National Natural Science Foundation of China (Grant Nos. 11874344, 61835004, 61327901, 11774335, 91536219, 11821404), Key Research Program of Frontier Sciences, CAS (No. QYZDY-SSW-SLH003), Anhui Initiative in Quantum Information Technologies (AHY020100, AHY060300), the Fundamental Research Funds for the Central Universities (Grant No. WK2030020019, WK2470000026), Science Foundation of the CAS (No. ZDRW-XH-2019-1).

$\textbf{Author Contributions}$
P.Y., Y.T. and W.-H.Z. contribute equally to this work.
P.Y. and G.C. planned and designed the experiment.
Y.T. and Y.M. proposed the framework of the theory and made the calculations.
P.Y. and W.-H.Z. carried out the experiment
assisted by Z.-Q.Z., J.-S.X., J.-S.T. and X.-Y.X., whereas X.-X.P. designed the computer
programs.
G.C. and Y.T. analyzed the experimental
results and wrote the manuscript assisted by Z.-Q.Y.
G.-C.G. and C-F.L. supervised
the project. All authors discussed the experimental
procedures and results.

\textbf{Competing interests:}
The authors declare no competing financial interests.

\clearpage{}
\end{document}